\documentclass[epjST]{svjour}
\usepackage{graphicx}
\usepackage{amsmath}

\usepackage{color}
\definecolor{azure}{rgb}{0.0, 0.5, 1.0}

\begin{document}
\title{Infrared Renormalons in Collider Processes}
\subtitle{}
\author{Silvia Ferrario Ravasio\inst{1}\fnmsep\thanks{\email{silvia.ferrarioravasio@physics.ox.ac.uk}}}
\institute{Rudolf Peierls Centre for Theoretical Physics,
  Clarendon Laboratory, Parks Road, Oxford OX1 3PU, UK}
\abstract{
Precise theoretical predictions are a key ingredient for an accurate determination 
of the structure of the Langrangian of particle physics, including its free parameters, which summarizes our understanding of the fundamental interactions among particles.
Furthermore, due to the absence of clear new-physics signals, precise theoretical calculations are required in order to pin down possible subtle deviations from the Standard Model predictions. 
The error associated with such calculations must be scrutinized, as non-perturbative power corrections, dubbed infrared renormalons, can limit the ultimate precision of truncated perturbative expansions in quantum chromodynamics. In this review we focus on linear power corrections that can arise in certain kinematic distributions relevant for collider phenomenology where an operator product expansion is missing, \emph{e.g.} those obtained from the top-quark decay products, shape observables and the transverse momentum of massive gauge bosons. Only the last one is found to be free from such corrections, while the mass of the system comprising the top decay products has a larger power correction if the perturbative expansion is expressed in terms of a short-distance mass instead of the pole mass.
A proper modelization of non-perturbative corrections is crucial in the context of shape observables to obtain reliable strong coupling constant extractions.
} 
\maketitle
\section{Introduction}
\label{intro}

The Standard Model~(SM) of particle physics encapsulates our understanding of the fundamental interactions among the elementary particles that constitute the building blocks of our universe.
Particle colliders like the Large Electron-Proton collider~(LEP) and the Large Hadron Collider~(LHC) enable us to probe such fundamental laws at unprecedented precision,
via comparison of accurate theoretical predictions and the experimental data.
Hower the SM is known to be incomplete, as it does not include a description of gravity, it cannot explain the origin of its fundamental parameters, neutrino oscillations, matter–antimatter asymmetry, the nature of dark matter and dark energy\ldots\,
Many Beyond SM extensions have been proposed but no clear signal of new physics has been observed yet. In this scenario, we expect new physics to manifest only via subtle deviations from SM predictions. 
Thus, in order to pin down such deviations and improve the accuracy of the determinations on the parameters of the SM Lagrangian, the experimental data must be confronted with precise theoretical calculations.

Calculations for collider processes are performed as a perturbative expansion in the strong coupling $\alpha_s$, where the theoretical uncertainty mainly arises from the truncation of this series.
However it is well-known that perturbative expansions in quantum field theory do not converge~\cite{Dyson:1952tj}, as a non-zero convergency radius will imply the validity of the theory also for negative values of the coupling constant, which instead prevents the existence of a stable vacuum.  
In particular, if we want to apply the method developed by Borel to resum factorially growing divergent series~\cite{Borel} to quantum chromodynamics~(QCD), we are forced to introduce ambiguous terms that lead to power corrections of the order $(\Lambda/Q)^p$, with $\Lambda$ an hadronic scale, $Q$ the hard-scattering scale and $p$ a positive integer. These power corrections are dubbed \emph{infrared renormalons} as they are associated with the growth of the strong coupling constant at small scales\footnote{Since the elecromagnetic coupling constant $\alpha_{em}$ increases at large energies, quantum electrodynamics is instead affected by ultraviolet renormalons.}.
The existence of an Operator Product Expansion~(OPE) provides a safe guideline for the classification of such power corrections, due to the correspondance between power-suppressed terms caused by the bad large-order behaviour of QCD and those associated from higher twists operators.
Thus, we can see that power corrections arise because perturbation theory is not
complete and we need to include non-perturbative contributions to recover the full result~\cite{Mueller:1993pa}.

Unfortunately, there is no OPE for the majority of the kinematic distributions measured at colliders, so we need to compute the perturbative coefficients at all orders to assess for example the presence of \emph{linear} power corrections ($\Lambda/Q$), which can lead to an ambiguity at the percent level for hard scales of the order of the $Z$ mass, possibly significantly limiting the ultimate uncertainty of theoretical predictions. 
 This can be done in the limit of an infinite and negative number of quark flavours $n_f$, where at the end one performs the replacement $n_f \to -6 \pi b_0$, where $b_0$ is the first coefficient of the QCD beta function 
\begin{equation}
b_0 = \frac{11 C_A}{12 \pi}-\frac{n_l}{6\pi},
\label{eq:b0}
\end{equation}
with $n_l$ being the real number of light flavours of the theory~\cite{Beneke:1994qe}. 
This procedure encapsulates an explicit dependence on the running of the strong coupling, which is the responsible of the factorial growth of the perturbative expansion, and for this reason it has been successfully employed in several contests, like the pole mass ambiguity~\cite{Ball:1995ni}, 
event shapes~\cite{Nason:1995np}, Drell-Yan~(DY) production~\cite{Beneke:1995pq}, fragmentation functions in $e^+e^-$ annihilation~\cite{Dasgupta:1996ki}. An exhaustive description of the phenomenological applications is discussed in Sec.~5 of Ref.~\cite{Beneke:1998ui}, which constitutes a comprehensive and detailed review on the topic of renormalons. In this contribution we instead only focus on most recent phenomenological applications.

In Sec.~\ref{sec:borel-divergent-qcd} we briefly summarize the Borel summation technique, emphasizing the residual ambiguity related to perturbative series in QCD. In Sec.~\ref{sec:large-nf} we introduce the large-$n_f$ limit to assess the presence of power corrections, and we discuss the application to the pole mass~(Sec.~\ref{sec:polemass}). In Sec.~\ref{sec:our large nf} we illustrate
how this method can be applied numerically for evaluating any arbitrary infrared-safe observable, following the proposal of Refs.~\cite{FerrarioRavasio:2018ubr,FerrarioRavasio:2020guj}. The application to the case of top-quark production and decay is summarized in Sec.~\ref{sec:top-prod}.
In Sec.~\ref{sec:Zpt} this method is employed to investigate the presence of linear renormalons in the transverse-momentum distribution of a gauge boson recoiling against a hard jet. 
In Sec.~\ref{sec:shape-obs} we discuss the evaluation of non-perturbative corrections for shape observables in $e^+e^-$ collisions.
Conclusions and outlooks are summarized in Sec.~\ref{sec:concl}.

\section{Factorially divergent series}
\label{sec:borel-divergent-qcd}
Let us consider a factorially-divergent series
\begin{equation}
\Sigma[\alpha] = \sum_{n=0}^\infty c_n \alpha^{n+1} \qquad \mbox{ with } \qquad c_n = (-b)^n n!, 
\label{eq:diff-sign}
\end{equation}
where $\alpha$ is the small (and positive) expansion parameter and $b$ is a real number. 
The smallest term of the series is reached for $|c_{n-1} \alpha^{n}|\approx |c_{n} \alpha^{n+1}|$, \emph{i.e.} for $n_{\rm min}\approx1/|b|\alpha$ and, if $n_{\min}\gg1$, has size
\begin{equation}
|c_{n_{\min}}\alpha^{n_{\min}+1}| \approx \frac{1}{|b|} \sqrt{\frac{2\pi}{n_{\rm min}}}{\rm e}^{-\frac{1}{\alpha|b|}}. \label{eq:minimum}
\end{equation}
Borel devised a summation method for alternating factorially-divergent series~\cite{Borel}, which allows us to rewrite eq.~\eqref{eq:diff-sign} for $b>0$ as
\begin{equation}
\Sigma[\alpha] = \int_0^{\infty} d\beta \,{\rm e}^{-\beta/\alpha} \sum_{n=0}^\infty \frac{c_n}{n!} \beta^n
=  \int_0^{\infty} d\beta \frac{{\rm e}^{-\beta/\alpha}}{1+\beta b}, \label{eq:borel-sum}
\end{equation}
which is well defined as there are no poles on the integration range. We immediately realise that if we want to use the Borel summation also when $b$ is negative, \emph{i.e.} if $b=-|b|$ and the coefficients in eq.~\eqref{eq:diff-sign} have all the same sign, there is a pole on the integration path: to interpret eq.~\eqref{eq:borel-sum} as estimate of the resummed expression we thus need to add a small imaginary part to the denominator of the integrand. The sign of this imaginary part is arbitrary, indeed we can define
\begin{equation}
\Sigma_{\pm}[\alpha]=\int_0^{\infty} d\beta \frac{{\rm e}^{-\beta/\alpha}}{1-\beta |b|\pm i\eta}. \label{eq:borel-sum}
\end{equation}
An estimate of the ambiguity of the resummed series is thus given by
\begin{equation}
\frac{\Sigma_+[\alpha]-\Sigma_-[\alpha]}{2\pi i} = \frac{e^{1/\alpha |b|}}{|b|}. \label{eq:ImPi}
\end{equation}
We see that this ambiguity is identical to the minimum term of the perturbative series given in eq.~\ref{eq:minimum}, modulo a normalization factor $\sqrt{2\pi/n_{\min}}$.

Let us now focus to QCD, where the expansion paramer is now the strong coupling $\alpha_s$ evaluated at a hard scale $Q$
\begin{equation}
\alpha_s \equiv \frac{1}{2 b_0 \ln\left(\frac{Q}{\Lambda_{\rm QCD}}\right)},
\end{equation}
where $b_0$ is given by eq.~\eqref{eq:b0} and $\Lambda_{\rm QCD}$ is the energy scale at which $\alpha_s$ becomes infinity. 
 The value of the strong coupling at another scale can be expressed as a function of $\alpha_s$:
\begin{equation}
\alpha_s(k)=\frac{\alpha_s}{1+ b_0 \alpha_s \ln \left(\frac{Q^2}{k^2} \right)} = \sum_{n=0}^\infty \left[  b_0 \ln \left(\frac{Q^2}{k^2} \right) \right]^n \alpha_s^{n+1}.
\end{equation}
When performing all-orders calculations, a large number of contributions can be resummed by evaluating the next-to-leading-order~(NLO) term at a running scale. This means that the contribution to an infrared-safe observable is approximatively given by
\begin{equation}
\!
 \int_0^{Q} \frac{d k}{k}\, k^{p} \,\alpha_s(k) = \sum_{n=0}^\infty \int_0^{Q} \frac{d k}{k}\, k^{p}\left[  b_0 \ln \left(\frac{Q^2}{k^2} \right) \right]^n\!\! \alpha_s^{n+1}=  Q^p \sum_{n=0}^{\infty} \left(\frac{2 b_0}{p}\right)^n \!\! n!\,\alpha_s^{n+1}, \label{eq:qcd}
\end{equation}
where $p$ is a positive integer as required from the infrared-safety property of the observable. Thus, since the first coefficient of the QCD $\beta$-function $b_0$ is positive, resumming soft emissions at all orders leads to same-sign factorially divergent series, with minimum term located at $n_{\min}=\frac{p}{2b_0\alpha_s}$. The smallest term of the series and the ambiguity on the Borel-resummed result are proportional to 
\begin{equation}
{\rm e}^{-\frac{p}{2 b_0 \alpha_s}}=(\Lambda_{\rm qcd}/Q)^p.
\label{eq:p}
\end{equation}
Thus we see that the nature of the power corrections induced by the bad large-order behaviour of QCD is encoded in the value of the parameter $p$ appearing in eq.~\eqref{eq:qcd}. The largest and most worrisome corrections arise for $p=1$ and are dubbed \emph{linear infrared renormalons}.

\section{The large-number-of-flavour approximation}
\label{sec:large-nf}
In absence of an OPE, we need to perform an all-order calculation to infer the value of  $p$ which appears in eq.~\eqref{eq:qcd}.
Higher-order corrections are accessible up to all orders in the coupling when considering the limit of a large number of flavours $n_f$, where the only contributions to be considered arise from the insertion of fermion loops in the gauge-field propagator, since they involve
powers of $n_f \alpha_s$. This implies that free gluon propagator must be replaced with the dressed one:
\begin{equation}
\frac{-i g^{\mu \nu}}{k^2-i\eta} \rightarrow\frac{-i g^{\mu \nu}}{k^2-i\eta} \frac{1}{1+\Pi(k^2,Q^2) }.
\end{equation}
The self-energy correction $\Pi(k^2,Q^2)$ is ultraviolet divergent, but usually no additional subtractions beyond those contained in the renormalized QCD Lagrangian are required, so we can evaluate $\Pi$ using the renormalized fermion-loop expression\footnote{When evaluating quantities which explicitly involve ultraviolet divergences, like the difference between the pole and the $\overline{\rm MS}$ mass, we need to compute $\Pi(k^2,Q^2)$ in  $D=4-2\epsilon$ dimensions (or using another regularization procedure) and  extract the finite part of the result only after having carried out all the other integrations.}
\begin{equation}
\Pi(k^2,Q^2)= -\frac{\alpha_s n_f}{6\pi}\left( \ln\frac{|k^2|}{Q^2}-i\pi \theta(k^2)+C \right),
\label{eq:Pi}
\end{equation}
where $C$ is a renormalization-scheme dependent constant, equal to $5/3$ in the $\overline{\rm MS}$ scheme. 
This procedure was originally proposed for quantum electrodynamics~(QED) \cite{Lautrup:1977hs}, but since the pre-factor of eq.~\eqref{eq:Pi} coincides with the fermionic contribution of the full QCD $\beta$-function~\eqref{eq:b0}, we can adopt this method also for QCD if at the end perform the replacement
\begin{equation}
n_f \rightarrow -6\pi b_0,
\label{eq:replacement}
\end{equation}
to recover the non-abelian property of the theory~\cite{Beneke:1994qe}. This naive non-abelianization is often referred as the large-$b_0$ approximation.

\subsection{Application: top-quark pole mass}
\label{sec:polemass}
To gauge the validity of this method with a concrete example, we consider the relation between the $\overline{\rm MS}$~($\bar{m}(\bar{m})$) and the pole mass~($m$) for the top quark. 
This relation is known up to order $\mathcal{O}(\alpha_s^4)$~\cite{Marquard:2015qpa}
\begin{align}
m = \bar{m}(\bar{m})\sum_{n=0}^{\infty}  c_i \alpha_s^i =& 163.643\left[1+ 0.42 \,\alpha_s+ 0.83 \,\alpha_s^2 + 2.4\, \alpha_s^3+\,8.5\,\alpha_s^4 \right]{\rm GeV}+\mathcal{O}(\alpha_s^{5})\nonumber 
\\=&163.643 + 7.557 + 1.617 + 0.501 + 0.197 \, 
{\rm GeV}+\mathcal{O}(\alpha_s^{5}).
 \label{eq:mpole}
\end{align}
From eq.~\eqref{eq:mpole} it is already evident that the coefficients $c_i$ are affected by a factorial growth.
This series can be computed at all-orders in the large-$b_0$ approximation, yielding~\cite{Ball:1995ni}
\begin{align}
m=&163.643 +7.557 + 2.345 + 0.584 + 0.241+ 0.127  \nonumber \\
&+ 0.085 + 0.067 + 0.063 + 0.067 \,{\rm GeV} +\mathcal{O}(\alpha_s^{10}).
\label{eq:mpoleb0}
\end{align}
Comparing eqs.~\eqref{eq:mpole} and~\eqref{eq:mpoleb0}, we can immediately notice that the $\mathcal{O}(\alpha_s^4)$ coefficient computed in this approximation is only 20\% 
larger then the exact one, as expected from the observation that the factorial growth is already visible from eq.~\eqref{eq:mpole}.

The asymptotic behaviour of the $c_i$ coefficients can also be inferred assuming the resummed series is renormalization-scale independent~\cite{Beneke:1994rs}
\begin{equation}
c_{n-1} \rightarrow N (2 b_0)^n \Gamma(1+n+b) \left(1+\sum_{k=1}^\infty \frac{s^k}{n^k}\right), \label{eq:cnas}
\end{equation}
where $b$ and $s_k$ can be expressed in terms of the full QCD $\beta$-function coefficients. Since $c_{n}\propto (2 b_0)^n$, we immediately notice a similarity with eq.~\ref{eq:qcd} with $p=1$, thus the pole-mass definition is affected by a linear renormalon.
The only unknown ingredient in eq.~\eqref{eq:cnas} is the overall normalization $N$. The authors of Ref.~\cite{Beneke:2016cbu} fitted the value of $N$ from the already-known perturbative coefficients and used the asymptotic formula in eq.~\eqref{eq:cnas} to estimated the missing higher-order contributions, finding
 \begin{align}
m=&163.643+0.577 + 1.617 + 0.501 + 0.195
+0.112 \nonumber \\
&+ 0.079 + 0.066 + 0.064 + 0.071 \,{\rm GeV} +\mathcal{O}(\alpha_s^{10}).
\label{eq:mpole-asym}
\end{align}
We can notice a remarkable agreement between the large-order terms computed in eq.~\eqref{eq:mpole-asym} and those obtained applying the large-$b_0$ approximation. Thus this confirm that resumming all-order corrections in $\alpha_s b_0$ provides us a simple but yet valid approximation to infer the main properties of the large-order behaviour of QCD.

We can use the smallest term of the series in eq.~\eqref{eq:mpole-asym} to gauge the ambiguity of the Borel-resummed series
\begin{equation}
\mbox{ambiguity} \approx \bar{m}(\bar{m})\,c_8\, \alpha_s^8\, \sqrt{\frac{8}{2\pi}} \approx 70~{\rm MeV},
\end{equation}
however this number is obtained neglecting charm- and bottom-quark mass effects. By including them, the estimated ambiguity becomes 110~MeV~\cite{Beneke:2016cbu}, while the approach adopted by the authors of Ref.~\cite{Hoang:2017btd} leads to 250~MeV. 
If the top pole-mass ambiguity is of the order of 100 or 200~MeV is not crucial yet, given that the most accurate LHC measurements reach 500~MeV error~\cite{Khachatryan:2015hba,Aaboud:2018zbu}. 
However, having a method that allows us to determine the nature of the non-perturbative correction for a generic infrared-safe observable can help us assessing how the pole-mass ambiguity propagates to top-dependent observables. We can also apply this method in the context of shape observables, where non-perturbative effects are large and need a proper modelization, or to kinematic distributions like the transverse momentum of gauge bosons, to determine the presence or the absence of linear $\Lambda$ terms.

\subsection{Practical implementation}
\label{sec:our large nf}
\begin{figure}[t!]
\includegraphics[width=\textwidth]{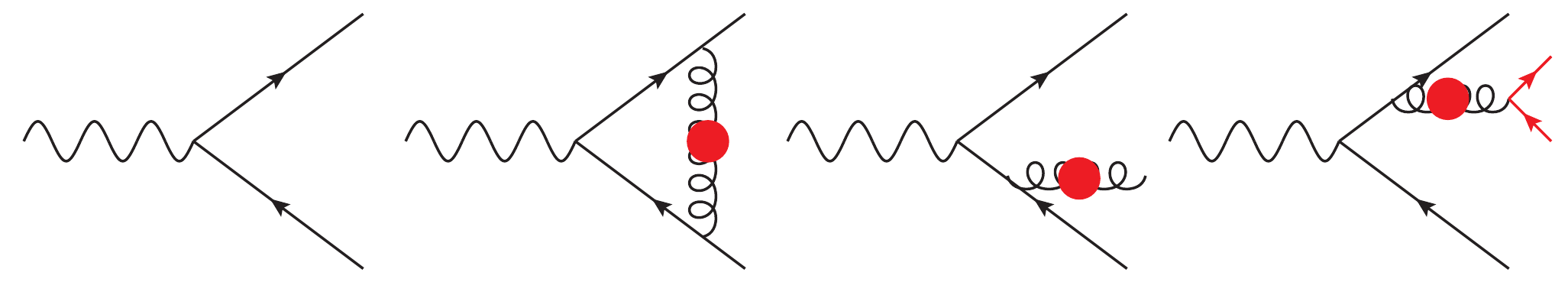}\\
$\phantom{a}$\hspace{1.35cm}(b)
\hspace{2.5cm}
(v) 
\hspace{2.5cm}
($g$) 
\hspace{2.5cm}
($\rm q\bar{q}$)
\label{fig:allnf}
\caption{A sample of the contributions that need to be included in order to
compute the leading large $b_0$ corrections to the process $\gamma^*\to d\bar{d}$.
The solid blob is defined via the recursive relation in eq.~\eqref{eq:blob}.}
\end{figure} 

In this section we describe how to compute all-orders corrections in the large-$b_0$ limit for a generic process which does not contain any gluon at the lowest perturbative order, following the recipe introduced in Ref.~\cite{FerrarioRavasio:2018ubr}.
The expression for a generic observable $O$ function of the final-state kinematics $\Phi$ for $n_f\to \infty$ is given by
\begin{align}
O= & \phantom{+}\int \frac{d\sigma_{\rm b}(\Phi_{\rm b})}{d\Phi_{\rm b}} d\Phi_{\rm b} O(\Phi_{\rm b}) + \int \frac{d\sigma_{\rm v}(\Phi_{\rm b})}{d\Phi_{\rm b}} d\Phi_{\rm b} O(\Phi_{\rm b}) \nonumber \\
&+ \int \frac{d\sigma_{g}(\Phi_{g})}{d\Phi_{g}} d\Phi_{g} O(\Phi_{g})+\int \frac{d\sigma_{q\bar{q}}(\Phi_{q\bar{q}})}{d\Phi_{q\bar{q}}} d\Phi_{q\bar{q}}\, O(\Phi_{q\bar{q}}),
\label{eq:Obsnf}
\end{align}
where $\sigma_{\rm b}$ corresponds to the Born cross section, $\sigma_{\rm v}$ corresponds to the virtual cross section, $\sigma_g$ comes from the emission of an onshell gluon and $\sigma_{q\bar{q}}$ is obtained from the emission of an offshell gluon which splits into a $q\bar{q}$ pair. Examples of the Feynman diagrams contributing to $\sigma_{\rm b}$, $\sigma_{\rm v}$, $\sigma_g$ and $\sigma_{q\bar{q}}$ are given in Fig.~\ref{fig:allnf}, where the solid red blob represents the insertion of the fermionic loops in all possible ways,
and can be defined with this recursive relation
\begin{equation}
\raisebox{-5mm}{
\includegraphics[width=0.85\textwidth]{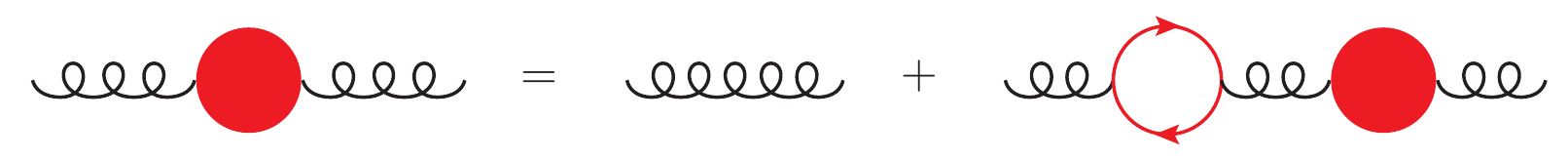}}.
\label{eq:blob}
\end{equation}
We stress that in the $\rm (q\bar{q})$ contribution of Fig.~\ref{fig:allnf} we can always assume that the $q\bar{q}$ pair arising from the gluon splitting has a different flavour from other quarks already present in the LO (b) diagram, as interference effects would be suppressed by one inverse power of $n_f$. 
We label with $\Phi_{\rm b}$ and $\Phi_g$ the phase space for the Born configuration and for the process containing an additional gluon $g$, while $\Phi_{q\bar{q}} $ is the phase space for the contribution with an additional $q\bar{q}$ pair, which can be conveniently rewritten as 
\begin{equation}
d\Phi_{q\bar{q}} = \frac{d m^2_{q\bar{q}}}{2\pi} d\Phi_{g^*}^{\scriptsize (m_{q\bar{q}})} d\Phi_{\rm dec},
\end{equation}
where ${g^*}$ denotes an offshell gluon with mass $m_{q\bar{q}}$ and $\Phi_{\rm dec}$ is the phase space for the $g^*\to q\bar{q}$ decay.

Manipulating eq.~\eqref{eq:Obsnf} and performing the replacement in eq.~\eqref{eq:replacement} we get~\cite{FerrarioRavasio:2018ubr} 
\begin{align}
O = O_{\rm b}-\int_0^{+\infty} d\lambda\,\frac{1}{\alpha_s(\mu)} \frac{\partial T(\lambda;\mu)}{\partial \lambda} \,\alpha_{\rm eff}(\lambda),
\label{eq:Ofin}
\end{align}
where $O_{\rm b}$ is the LO prediction,
\begin{equation}
\alpha_{\rm eff}(\lambda ) =  \frac{\arctan(\pi \,b_0\, \alpha_s(\lambda \,{\rm e}^{-\frac{C}{2}}))}{\pi b_0}= \frac{1}{\pi b_0}\arctan\!\! \left[\frac{\pi b_0 \alpha_s\!(\mu)}{1+b_0\, \alpha_s\!(\mu) \ln\! \left(\!\frac{\lambda^2}{\mu^2 {\rm e}^{C}}\!\right)} \right]\!\!,
\label{eq:effcoupl}
\end{equation}
with $C$ a renormalization-scheme dependent constant that we choose to be equal to
\begin{equation}
C = \frac{1}{b_0}\left[ \left(\frac{67}{18}-\frac{\pi^2}{6}\right) C_A -\frac{5}{3}\frac{n_l}{6\pi}\right],
\end{equation}
so that $\alpha_s(\lambda \,{\rm e}^{-C/2})$ corresponds to $\alpha_s(\lambda)$ in the Catani-Marchesini-Webber~(CMW) scheme~\cite{Catani:1990rr}, and
\begin{align}
T(\lambda;\mu) =& \int \frac{d\sigma^{(1)}_{\rm v}(\Phi_{\rm b};\lambda)}{d\Phi_{\rm b}} d\Phi_{\rm b} O(\Phi_{\rm b}) \nonumber + \int \frac{d\sigma_{g^*}^{(1)}(\Phi_{g^*}^{(\lambda)})}{d\Phi_{g^*}^{(\lambda)}} d\Phi_{g^*}^{(\lambda)} O(\Phi_{g^*}^{(\lambda)})
\nonumber \\
& 
+\frac{6\pi \lambda^2}{n_f \alpha_s(\mu)}\int \frac{d\sigma^{(2)}_{q\bar{q}}(\Phi_{q\bar{q}})}{d\Phi_{q\bar{q}}} 
\delta(m_{q\bar{q}}^2-\lambda^2)\,
d\Phi_{q\bar{q}}  \left[O(\Phi_{q\bar{q}})-O(\Phi_{g^*}^{(\lambda)})\right],
\label{eq:Tlambda}
\end{align}
where $\sigma^{(1)}_{\rm v}$ and $\sigma^{(1)}_{g^*}$ are the $\mathcal{O}(\alpha_s)$ corrections to the LO cross section, computed with a gluon of mass $\lambda$, while $\sigma_{q\bar{q}}^{(2)}$ corresponds to the ($\rm q \bar{q}$) contribution of Fig.~\ref{fig:allnf} without any blob insertion in the offshell gluon propagator.
The $\mu$ dependence cancels in the ratio $T(\lambda; \mu)/\alpha_s(\mu)$, so that $O$ is effectively $\mu$ independent.
For large $\lambda$, $T(\lambda;\mu) \to \frac{1}{\lambda^2}$ so the integral in eq.~\eqref{eq:Ofin} is UV convergent. For vanishing $\lambda$
\begin{equation}
\lim_{\lambda \to 0} T(\lambda; \mu) = \int \frac{d\sigma^{(1)}_{\rm v}(\Phi_{\rm b};0)}{d\Phi_{\rm b}} d\Phi_{\rm b} O(\Phi_{\rm b})  + \int \frac{d\sigma_{g}^{(1)}(\Phi_{g})}{d\Phi_{g}} d\Phi_{g} O(\Phi_{g}),
\label{eq:NLO}
\end{equation}
which corresponds to the usual NLO corrections computed using a mass regularization for processes which do not contain incoming partons. If we consider a process with an incoming quark leg, eq.~\eqref{eq:NLO} contains divergent $\ln\lambda$ terms due to the mis-cancellation of initial-state collinear singularities. We thus need to include to $T(\lambda; \mu)$ the term~\cite{FerrarioRavasio:2020guj}
\begin{equation}
\Delta T(\lambda; \mu) = -\int d{\Phi}_{\rm b} \frac{d\sigma_{\rm b}(xQ)}{d \Phi_{\rm b}} \int_x^1 \frac{dz}{z}\frac{f_q(x/z, \mu_F)}{f_q(x, \mu_F)} C^{(1)}_{q\bar{q}}(\lambda/\mu_F, z, \mu),
\label{eq:TISR}
\end{equation}
where $Q$ is the momentum of the incoming proton, while $x$ is the energy fraction of the quark which enters the LO hard scattering process, $\mu_F$ is the factorization scale\footnote{The $\mu_F$ dependence cancels in the integral if the PDF evolution is also computed in the large-$b_0$ approximation.}, $f_q$ is the parton distribution function~(PDF) for the incoming quark and $C^{(1)}_{q\bar{q}}$ corresponds to the $\mathcal{O}(\alpha_s)$ correction to the Deep Inelastic Scattering~(DIS) cross section, computed with a massive gluon, normalised to the LO prediction, stripping all the PDF factors~\cite{Beneke:1995pq}.
In this way $T(\lambda;\mu)$ smoothly approaches the conventional NLO result where the PDF has been renormalized in the DIS scheme.

We notice that if 
\begin{equation}
T^\prime(\lambda;\mu)=\left.\frac{\partial T(\lambda;\mu)}{\partial \lambda} \right|_{\lambda=0}=-\alpha_s(\mu) A \neq 0,
\label{eq:deriv}
\end{equation}
the small-$\lambda$ contribution to~\eqref{eq:Ofin} gives rise to the following perturbative series
\begin{equation}
O \sim A \mu{\rm e}^{\frac{C}{2}} \sum_{n=0}^{\infty} (2 b_0)^n \,n!\, \alpha_s^{n+1},
\end{equation}
which has minimum equal to $A  {\rm e}^{\frac{C}{2}}\sqrt{\frac{\alpha_s \pi}{b_0}} \Lambda$, or equivalently leads to a Borel-resummed result with ambiguity $\frac{A  {\rm e}^{\frac{C}{2}}}{2 b_0} \Lambda$. 
Thus, a linear-$\lambda$ in $T(\lambda)$ leads to linear renormalons, while if the derivative in \eqref{eq:deriv} is zero, only higher power corrections will appear.
The term included in eq.~\ref{eq:TISR} does not include any linear $\lambda$-dependence, as the form factor which gives the DIS cross section obeys an OPE where power corrections are controlled by higher twist operators, and the
dominant power corrections, corresponding to twist 4, are quadratic.

The ingredients appearing in eq.~\eqref{eq:Ofin} are not new, indeed
an effective coupling $\alpha_{\rm eff}(\lambda)$ similar to the one in eq.~\eqref{eq:effcoupl} naturally arises when applying dispersion relations to bubble diagrams appearing in the virtual diagrams (see Ref.~\cite{Beneke:1998ui} and references therein). Furthermore infrared renormalons in shape observables were typically inferred 
convoluting radiative corrections due to soft emissions with an
 effective coupling~\cite{Dokshitzer:1995zt}. The first line of eq.~\eqref{eq:Tlambda} corresponds to the NLO calculation performed using a gluon of mass $\lambda$, as a matter of fact  linear power corrections were originally inferred using the ``massive gluon method'', which was used for example by the authors of Ref.~\cite{Beneke:1995pq} to show that the DY cross section is not affected by linear renormalons. 
The second line of eq.~\eqref{eq:Tlambda}
is identical to zero for inclusive observables insensitive to the gluon splitting appearing in the contribution $(\rm q\bar{q})$ of Fig.~\ref{fig:allnf}, but it must be included for observables sensitive to the kinematics of the whole final state, as already shown in Ref.~\cite{Nason:1995np}. Eq.~\eqref{eq:Ofin} brings all these ingredients together and allows us to compute results in the large-$b_0$ approximation exactly,
using a combination of analytic and numerical methods, in such a way that \emph{any} arbitrary process, which does not contain gluons at LO, can be computed.

\section{All-order corrections for top-mass determinations}
\label{sec:top-prod}

The mass of the top quark is one key parameters of the standard model, as it is important non only for the top phenomenology, but also because it largely affects many other parameters (\emph{e.g.} gauge boson masses, Higgs trilinear coupling). As already discussed in Sec.~\ref{sec:polemass}, the pole mass definition is affected by a linear renormalon and hence has an ambiguity of roughly 100--200 MeV~\cite{Beneke:2016cbu,Hoang:2017btd}.
In Ref.~\cite{FerrarioRavasio:2018ubr}
we investigated the asymptotic behaviour of kinematic distributions of the top decay-products computed in terms of the
pole mass and of the $\overline{ \rm MS}$ mass (that we can consider as a proxy of all the short-distance mass schemes)
in a simplified theoretical framework where we understand some aspects concerning the non-perturbative corrections to the pole mass.
To achieve this task, formula  \eqref{eq:Ofin} was applied to the process of single-top production and decay, \emph{i.e.} $W^*\to t \bar{b} \to W b\bar{b}$. The calculation of $T(\lambda;\mu)$ was performed in the complex pole scheme, with $\mu={\rm Re}(m_t)$, in order to include finite top-width~($\Gamma_t$) effects and interferences between radiative corrections in the production process $W^*\to t \bar{b}$, and in the top decay $t\to W b$. 
We stress that eq.~\eqref{eq:Ofin} cannot be evaluated directly in four dimensions if the top mass is renormalized in the $\overline{\rm MS}$ scheme, as $T(\lambda; \mu)$ would still contain ultraviolet divergencies. Details on the computation of $T(\lambda; \mu)$ in the $\overline{\rm MS}$ scheme can be found in Sec.~3.3 of Ref.~\cite{FerrarioRavasio:2018ubr}, however if we are only interested in the small-$\lambda$ behaviour we can write
\begin{equation}
T^\prime_{\rm \overline{MS}}(\lambda; \mu) \approx T^\prime_{\rm pole}(\lambda; \mu)-\alpha_s(\mu) \frac{C_F}{2} \frac{\partial O_{\rm b}(m_t, m_t^*)}{\partial {\rm Re}(m_t)},
\label{eq:counterterm}
\end{equation}
where the $-\alpha_s(\mu) \frac{C_F}{2}$ coefficient arises from the computation of the one-loop top self-energy correction with a massive gluon.

When computing the total cross section, the linear-$\lambda$ sensitivity is given only by the insertion of the pole-mass conterterm, which cancels exactly with the extra term in eq.~\eqref{eq:counterterm} when using the $\overline{\rm MS}$ scheme.
This cancellation takes place irrespectively of the value of $\Gamma_t$, and hence we can affirm that the total cross section in the $\overline{\rm MS}$ scheme is always free from linear renormalons, even if the calculation is performed in the narrow-width approximation~(NWA).
The smallest term of the series in the pole scheme is reached for $n_{\min} =8 \approx 1/(2 b_0 \alpha_s(m_t))$.

\begin{figure}[t!]
\includegraphics[height=0.24\textheight]{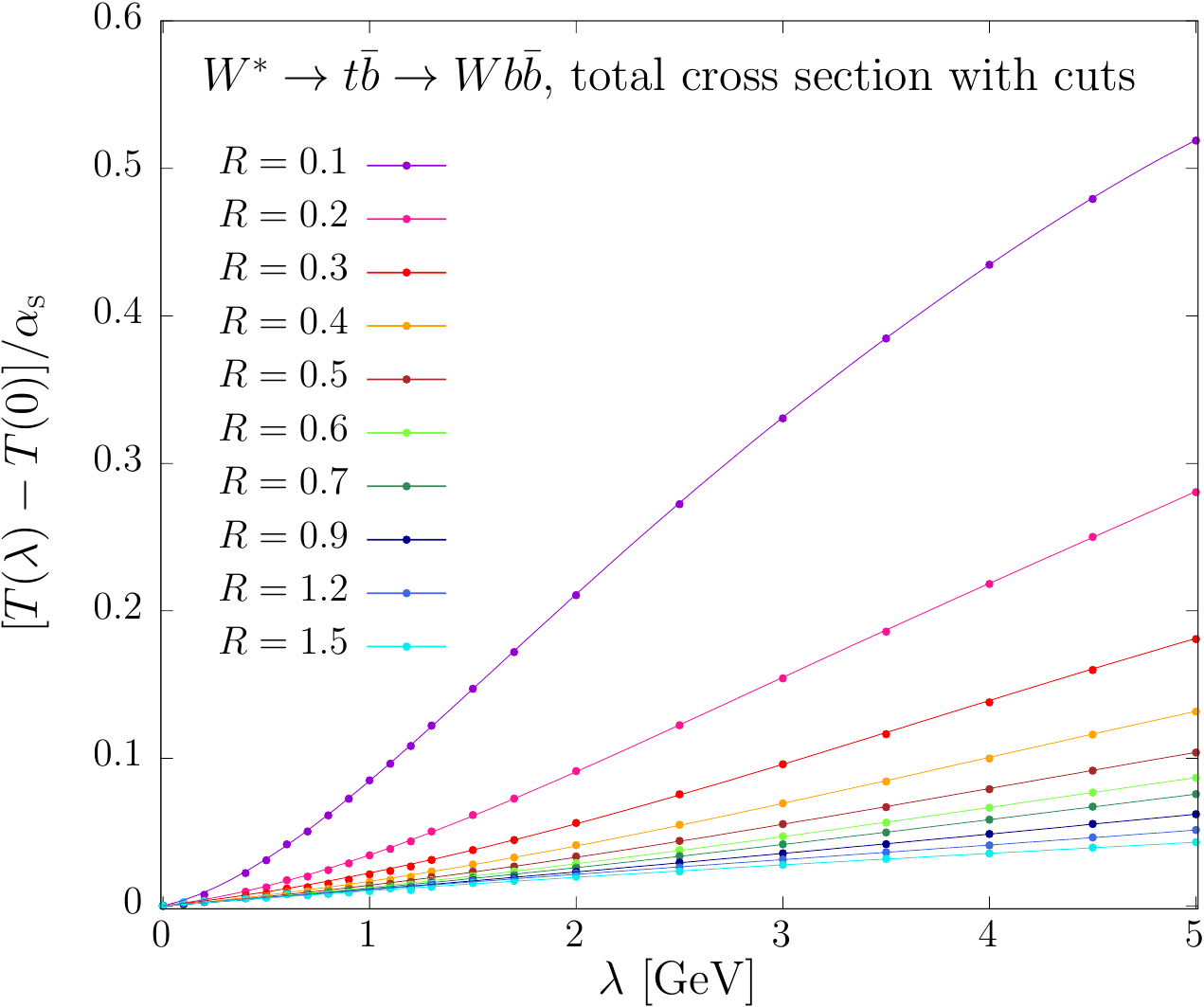}\hfill
\includegraphics[height=0.24\textheight]{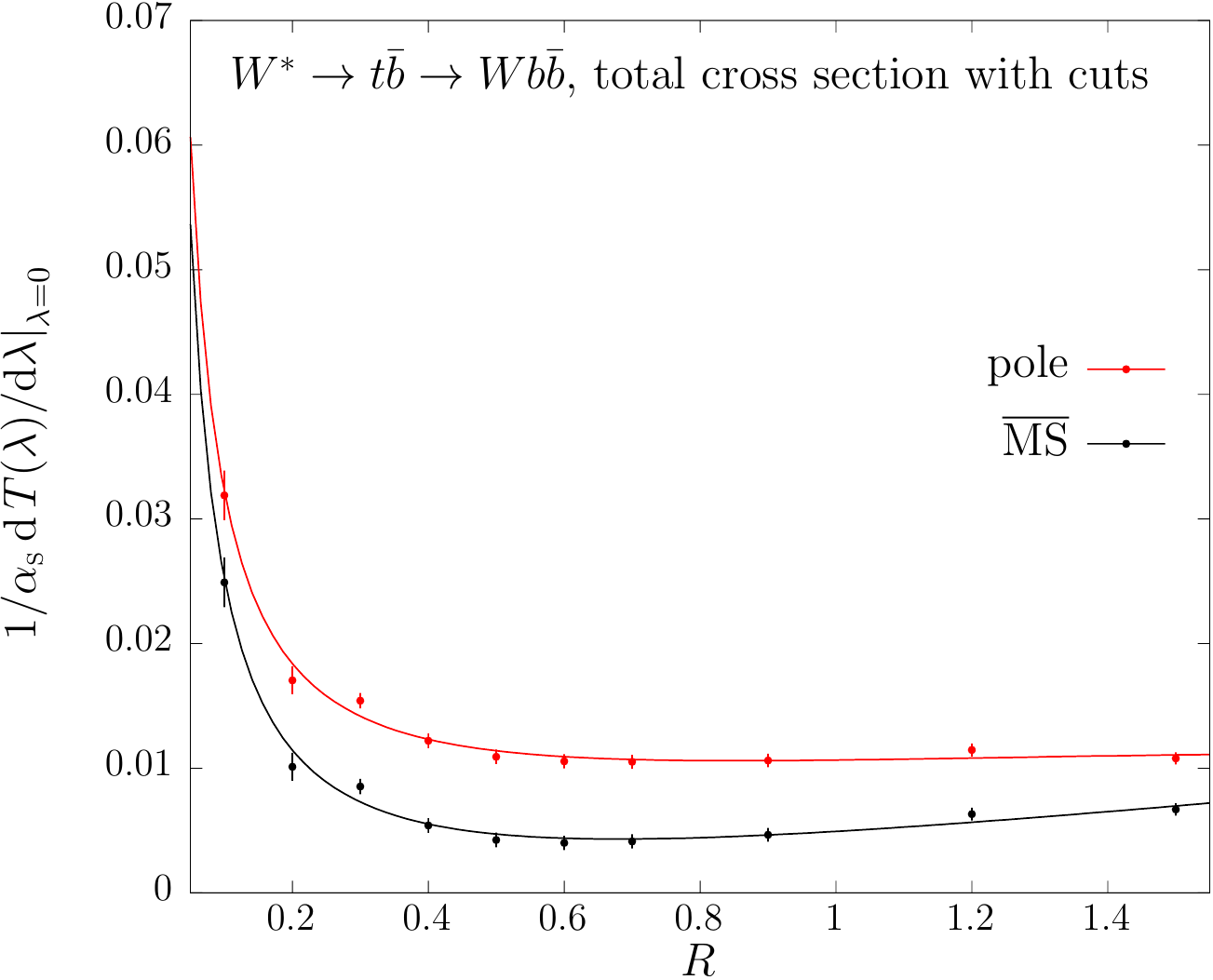}
\caption{Left panel: $T(\lambda)$ function, computed in the pole-mass scheme, for the fiducial cross section for $W^* \to b \bar{b} W$ requiring two resolved $b$-jets for several value of the jet radius $R$.
Right panel: $T^\prime(0)$ as a function of $R$  in the pole~(red) and in the $\overline{\rm MS}$ scheme. }
\label{fig:sigcut}
\end{figure}
If we want to define a fiducial cross section, requiring for example the presence of two different $b$ and $\bar{b}$ jets, we introduce a linear-$\lambda$ slope which diverges as $1/R$, where $R$ is the jet radius, irrespectively of the jet algorithm and of the mass scheme used, as we become sensitive to jet-related renormalons~\cite{Dasgupta:2007wa}.
This is illustrated in Fig.~\ref{fig:sigcut}, where we show the function $T(\lambda)$ computed for different values of the jet radius $R$ in the pole scheme (left panel) and its derivative for $\lambda=0$, both in the pole and in the $\overline{\rm MS}$ scheme.
 If we choose $R=1.5\approx\pi/2$, the slope in the $\overline{\rm MS}$ scheme is much smaller than in the pole scheme, thus yielding an improved perturbative convergence.

Interestingly, leptonic observables (in the collision frame) expressed in terms of the $\overline{\rm MS}$ mass are linear-renormalon free only if a non-zero $\Gamma_t$ value is adopted\footnote{Leptonic observables computed in the top frame are free from linear renormalons once the $\overline{\rm MS}$ mass is used because an OPE forbids such power corrections.}.
Indeed the presence of a finite $\Gamma_t=1.33$~GeV modifies the behaviour
of the integral in eq.~\ref{eq:Ofin} for scales smaller than $\Gamma_t$, preventing the top from being onshell. If we expand the effective coupling~\eqref{eq:effcoupl} in series of $\alpha_s(\mu)$, with $\mu={\rm Re}(m_t)=172.5$~GeV, we realise that $\lambda$ values larger than $ \Gamma_t$ give the dominant corrections till order $n\approx 1+\ln(\mu/\Gamma_t)\approx6$, so we need to compute many perturbative orders before realizing the improved perturbative convergence of the $\overline{\rm MS}$ with respect to the pole scheme. This is illustrated also in Tab.~\ref{tab:Ew}, where the coefficients of the perturbative expansion
\begin{equation}
\langle E_W \rangle = \sum_{i=0}^\infty c_i \alpha_s^i
\end{equation}
are shown.
\begin{table}
\centering
\begin{tabular}{c|c|c|c|c|}
  \cline{2-5}
  & \multicolumn{4}{|c|}{ $\phantom{\Big|}$ $\displaystyle \langle E_W
    \rangle$ $\phantom{\Big|}$ [GeV]}
 \\ \cline{2-5}
  & \multicolumn{2}{|c|}{ \phantom{\Big|}pole scheme \phantom{\Big|}}& \multicolumn{2}{|c|}{$\overline{\rm MS}$ scheme}
 \\ \cline{1-5}
 \multicolumn{1}{|c|}{$\phantom{\Big|} i \phantom{\Big|}$}  & $c_i $ & $ c_i\,\alpha_s^i$  & $c_i $ & $ c_i \, \alpha_s^i$ 
 \\ \cline{1-5}
 \multicolumn{1}{|c|}{ $\phantom{\Big|}$ 0 $\phantom{\Big|}$}& $121.5818$ & $121.5818$& $120.8654$ & $120.8654$
 \\ \cline{1-5}
 \multicolumn{1}{|c|}{$\phantom{\Big|}$           1 $\phantom{\Big|}$}& $-1.435 \, (0) \times 10^{    1}$ & $-1.552 \, (0) \times 10^{    0}$ & $-7.192 \, (0) \times 10^{    0}$ & $-7.779 \, (0) \times 10^{   -1}$ 
 \\ \cline{1-5}
 \multicolumn{1}{|c|}{$\phantom{\Big|}$           2 $\phantom{\Big|}$}& $-4.97 \, (4) \times 10^{    1}$ & $-5.82 \, (4) \times 10^{   -1}$ & $-3.88 \, (4) \times 10^{    1}$ & $-4.54 \, (4) \times 10^{   -1}$ 
 \\ \cline{1-5}
 \multicolumn{1}{|c|}{$\phantom{\Big|}$           3 $\phantom{\Big|}$}& $-1.79 \, (5) \times 10^{    2}$ & $-2.26 \, (6) \times 10^{   -1}$ & $-1.45 \, (5) \times 10^{    2}$ & $-1.84 \, (6) \times 10^{   -1}$ 
 \\ \cline{1-5}
 \multicolumn{1}{|c|}{$\phantom{\Big|}$           4 $\phantom{\Big|}$}& $-6.9 \, (4) \times 10^{    2}$ & $-9.4 \, (6) \times 10^{   -2}$ & $-5.7 \, (4) \times 10^{    2}$ & $-7.8 \, (6) \times 10^{   -2}$ 
 \\ \cline{1-5}
 \multicolumn{1}{|c|}{$\phantom{\Big|}$           5 $\phantom{\Big|}$}& $-2.9 \, (3) \times 10^{    3}$ & $-4.4 \, (5) \times 10^{   -2}$ & $-2.4 \, (3) \times 10^{    3}$ & $-3.5 \, (5) \times 10^{   -2}$ 
 \\ \cline{1-5}
 \multicolumn{1}{|c|}{$\phantom{\Big|}$           6 $\phantom{\Big|}$}& $-1.4 \, (3) \times 10^{    4}$ & $-2.2 \, (4) \times 10^{   -2}$ & $-1.0 \, (3) \times 10^{    4}$ & $-1.7 \, (4) \times 10^{   -2}$ 
 \\ \cline{1-5}
 \multicolumn{1}{|c|}{$\phantom{\Big|}$           7 $\phantom{\Big|}$}& $-8 \, (2) \times 10^{    4}$ & $-1.3 \, (4) \times 10^{   -2}$ & $-5 \, (2) \times 10^{    4}$ & $-8 \, (4) \times 10^{   -3}$ 
 \\ \cline{1-5}
 \multicolumn{1}{|c|}{$\phantom{\Big|}$           8 $\phantom{\Big|}$}& $-5 \, (2) \times 10^{    5}$ & $-9 \, (4) \times 10^{   -3}$ & $-2 \, (2) \times 10^{    5}$ & $-4 \, (4) \times 10^{   -3}$ 
 \\ \cline{1-5}
 \multicolumn{1}{|c|}{$\phantom{\Big|}$           9 $\phantom{\Big|}$}& $-3 \, (2) \times 10^{    6}$ & $-7 \, (4) \times 10^{   -3}$ & $-1 \, (2) \times 10^{    6}$ & $-2 \, (4) \times 10^{   -3}$ 
 \\ \cline{1-5}
 \multicolumn{1}{|c|}{$\phantom{\Big|}$          10 $\phantom{\Big|}$}& $-3 \, (2) \times 10^{    7}$ & $-6 \, (5) \times 10^{   -3}$ & $0 \, (2) \times 10^{    6}$ & $-1 \, (5) \times 10^{   -4}$ 
 \\ \cline{1-5}
 \multicolumn{1}{|c|}{$\phantom{\Big|}$          11 $\phantom{\Big|}$}& $-3 \, (3) \times 10^{    8}$ & $-7 \, (6) \times 10^{   -3}$ & $0 \, (3) \times 10^{    6}$ & $0 \, (6) \times 10^{   -5}$ 
 \\ \cline{1-5}
 \multicolumn{1}{|c|}{$\phantom{\Big|}$          12 $\phantom{\Big|}$}& $-4 \, (3) \times 10^{    9}$ & $-9 \, (9) \times 10^{   -3}$ & $0 \, (3) \times 10^{    8}$ & $1 \, (9) \times 10^{   -3}$ 
 \\ \cline{1-5}
 \end{tabular}
 \caption{Perturbative expansion of the average value of the $W$ boson energy in $W^*\to W b\bar{b}$, where the top mass is computed in the pole scheme~(left) and in the $\overline{\rm MS}$ scheme~(right). Statistical uncertainty is reported in parenthesis.}
 \label{tab:Ew}
\end{table}
Furthermore leptonic observables have a reduced top-mass sensitivity compared to other observables traditionally used to infer $m_t$ like, for example, the mass of the system comprising the top decay products~$M_{Wb_j}$ . For example, for $E=300$~GeV, $m_t=172.5$~GeV and $m_W=80.4$~GeV we have that at LO
\begin{equation}
\frac{\partial \langle E_W \rangle}{\partial m_t} \approx 0.1, \qquad \frac{\partial \langle M_{Wb_j} \rangle}{\partial m_t} \approx 1.
\end{equation}
This means that an uncertainty $\delta E_W$ on the energy of the $W$ boson will lead to an uncertainty of $10 \delta E_W$ on the value of the top mass $m_t$.
 
\begin{figure}[t!]
\includegraphics[height=0.24\textheight]{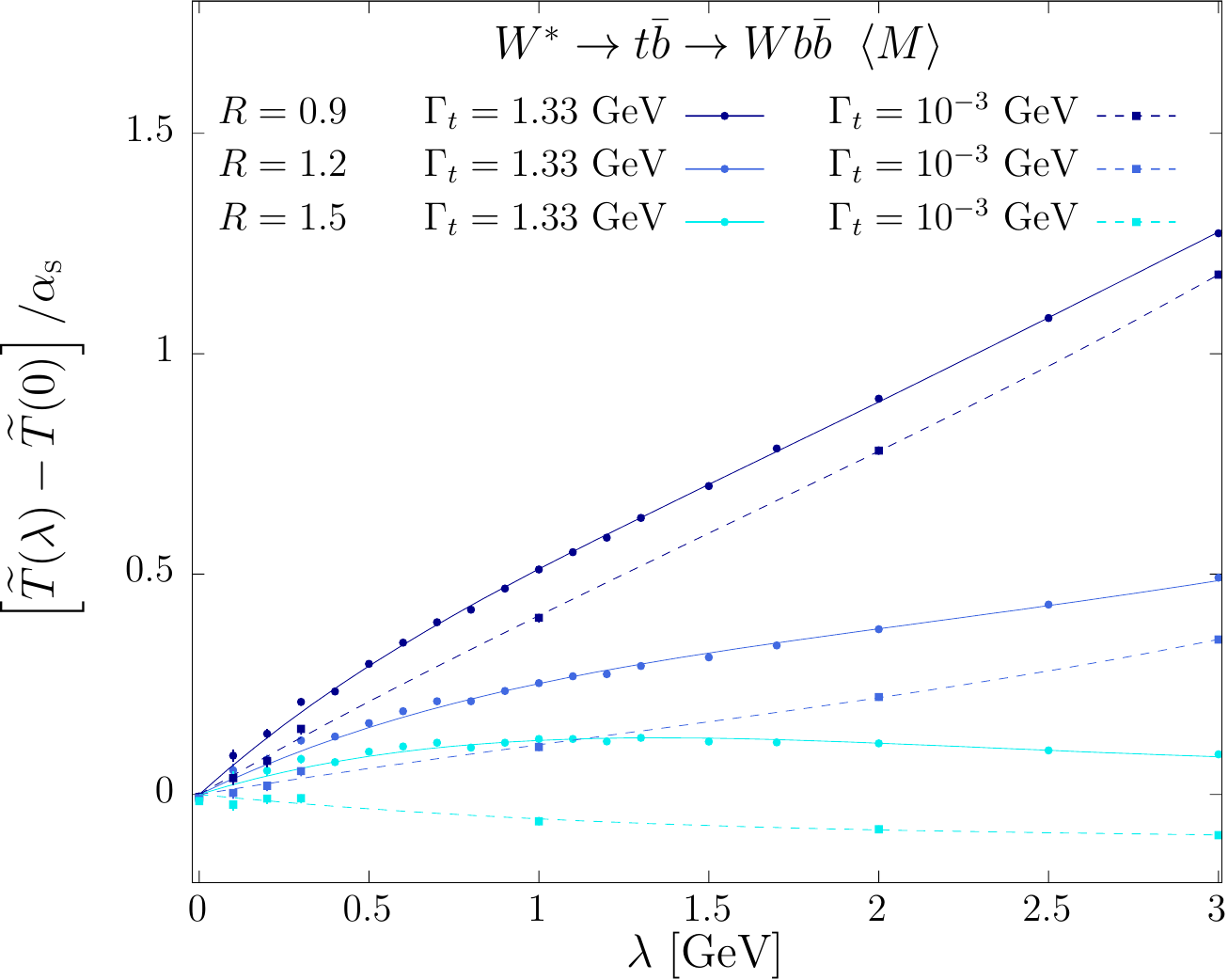}\hfill
\includegraphics[height=0.24\textheight]{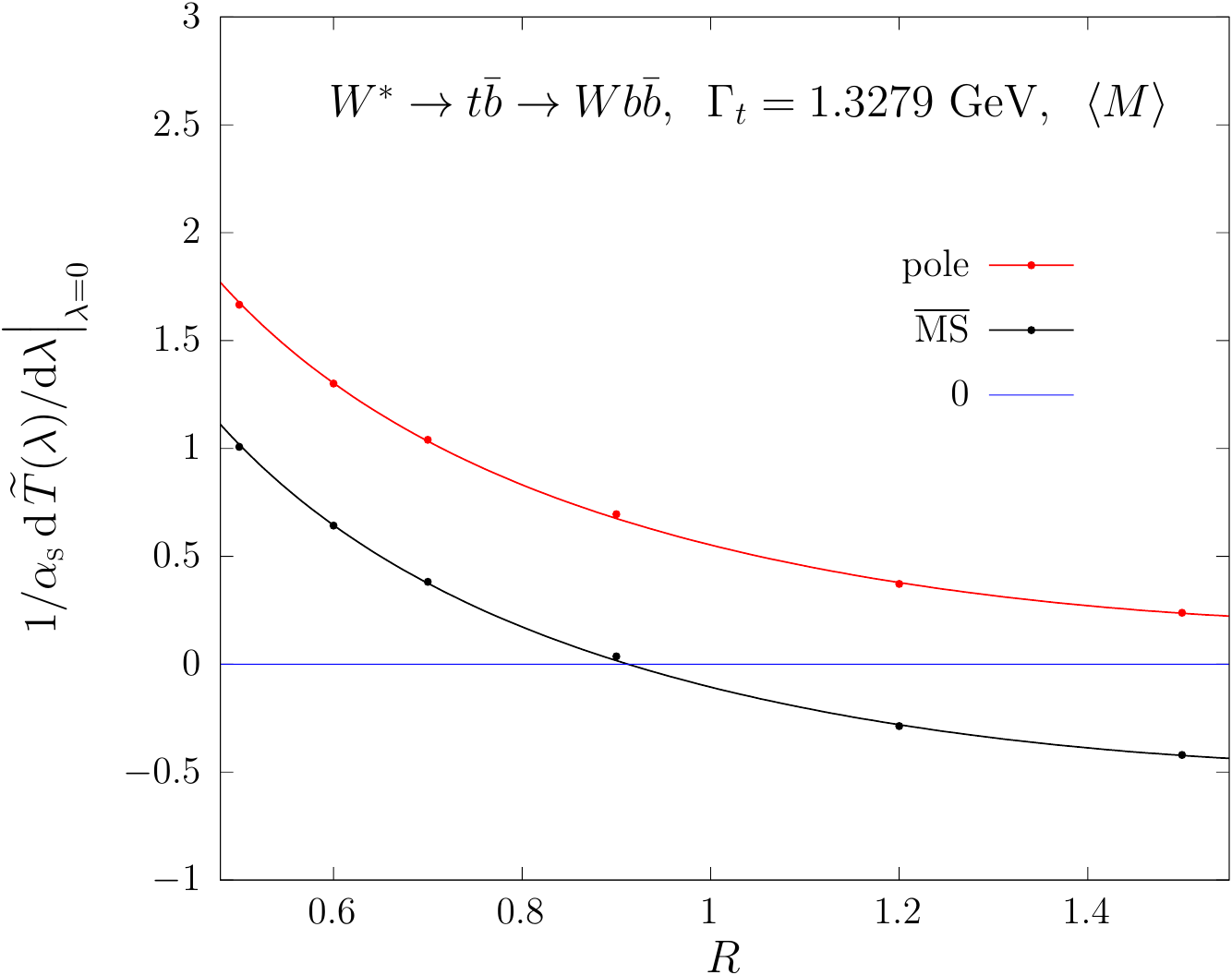}
\caption{Left panel: $T(\lambda)$ function, computed in the pole-mass scheme, for the mass of the $W$--b--jet, for several value of the jet radius $R$, for a vanishing top width~(dashed) and for $\Gamma_t=1.3279$~GeV.
Right panel: $T^\prime(0)$ as a function of $R$  in the pole~(red) and in the $\overline{\rm MS}$ scheme for $\Gamma_t=1.3279$~GeV. }
\label{fig:Mrec}
\end{figure}

For this reason, we investigate the linear-$\lambda$ sensitivity of the $T(\lambda;\mu)$ function associated with the average value of the mass of the $W$-boson and $b$-jet system $M_{W b_{\rm jet}}$. If we choose $R=1.5$ and we perform the calculation in NWA, we notice that $T^\prime_{\rm pole}(\lambda; \mu) \approx 0$, so there are no linear power-corrections. This is due to the fact that if $\Gamma_t=0$ and we are inclusive enough in the definition of the $b$ jet, the mass of the system comprising the top decay products is identical to the top pole-mass, which is not displaced by any radiative corrections by definition. This means $\langle M_{W b_{\rm jet}}\rangle = m_{\rm pole}$, which implies that the perturbative series in the pole scheme is well defined and all the $\alpha_s^n$ corrections are 0 for any $n>0$. If instead we use the ${\rm \overline{MS}}$ scheme, we find the same factorially-divergent series that relates the pole and the  ${\rm \overline{MS}}$ mass. This means that the observable contains an intrinsic physical renormalon which cancels with the pole-mass one. As already discussed for the case of leptonic observables, the presence of a finite top-width $\Gamma_t$ modifies the large-order behaviour of the series, as the interference between the radiation in the production process and the top decay spoils this cancellation. This is shown in the left panel of Fig.~\ref{fig:Mrec}.
However, we still find that for $R=1.5$ using the pole scheme yields a smaller ambiguity as the power correction  in the ${\rm \overline{MS}}$ scheme is roughly three times larger, as illustrated in the right panel of Fig.~\ref{fig:Mrec}.

\section{Transverse momentum of massive gauge bosons}
\label{sec:Zpt}

The transverse momentum of the $Z$-boson $p_{T}^{\small Z}$ is one of the most precise observables measured at the LHC, due to its large yield and its clear leptonic signature,
and the experimental error associated with the normalised distribution is at
sub-percent level in the low-intermediate values of the transverse
momentum~\cite{Sirunyan:2019bzr,Aad:2019wmn}. The current
state of the art for the theoretical prediction is given by NNLO+N${}^3$LL${}^\prime$~\cite{Re:2021con}, where the error is instead at the percent level.
Resummation effects have been proven to have a large effect in the
region where the $Z$ boson transverse momentum is small.
Non-perturbative corrections associated with the low-$p_T$ region have been extensively 
studied~\cite{Dokshitzer:1978yd,Parisi:1979se,Collins:1984kg,Scimemi:2016ffw}, and no linear power-correction was found. This has a rather intuitive explanation since in this context they can be modelled as a gaussian transverse momentum smearing.  
Since this smearing is azimuthally symmetric, its first-order effects
cancel out, leaving only quadratic corrections.

\begin{figure}[t]
\includegraphics[width=0.45\textwidth]{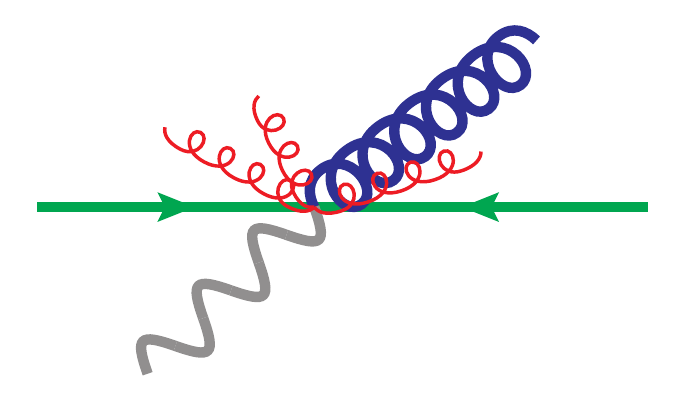} \hfill
\includegraphics[width=0.45\textwidth]{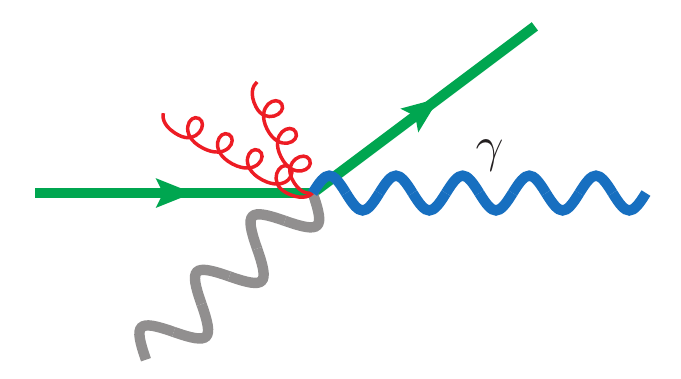}\\
$\phantom{a}$ \hspace{2.3cm}(a) \hspace{6.5cm}(b) 
\caption{Soft radiation pattern for $q\bar{q}\to Zg$~(a) and $q\gamma\to Zg$~(b) production. The green  solid lines represent quark or anti-quark lines, while the blue wiggly line represents the hard gluon against which the $Z$ boson~(gray wavy line) is recoiling against. 
 Soft gluon emissions are illustrated with red wiggly lines.}
\label{fig:Zpt}
\end{figure}

For moderate values of $p_{T}^{\small Z}$, the $Z$ boson recoils against a coloured final-state particle, and reliable theoretical predictions can be obtained from the calculation of $Z$+jet production at NNLO~\cite{Boughezal:2015ded,Ridder:2015dxa}. The presence of a linear correction of the order $\Lambda/p_{T}^{\small Z}$ is particularly worrisome as it can lead to a non-perturbative correction comparable to the current theoretical uncertainty and pose a serious limit on the ultimate uncertainty one can aim at with a pure perturbative calculation. Conversely to the inclusive DY case, when the $Z$ boson is produced in association with a hard jet, the soft-radiation pattern is not symmetric and this azimuthally asymmetry can in principle prevent the cancellation of linear power-corrections. This is illustrated in the left panel of Fig.~\ref{fig:Zpt}, which portrays the process $q\bar{q}\to Zg$, where additional soft gluons can only be emitted from two initial-final dipoles. 

Unfortunately, due to the presence of a gluon at the lowest perturbative order, such process cannot be computed using eq.~\eqref{eq:Ofin}. For this reason in Ref.~\cite{FerrarioRavasio:2020guj} we considered instead $q\gamma\to Zq$ (right panel of Fig.~\ref{fig:Zpt}), which still features the presence of an initial-final dipole and thus an azimuthally-asymmetric radiation pattern. QED divergences are removed using the FKS subtraction implemented in {\tt POWHEG BOX} framework~\cite{Frixione:2007vw,Alioli:2010xd,Jezo:2015aia} that we use for the numerical evaluation of $T(\lambda;\mu)$.
We compute the total cross section imposing a minimum $p_{T}^{Z}$ cut and we do not find any numerical evidence of linear-$\lambda$ terms. 
We do not find linear renormalons also for the $Z$-rapidity~$y_Z$ distribution computed with such $p_{T}^{Z}$ cut, as already found for the more inclusive DY production~\cite{Dasgupta:1999zm}.
This is illustrated in Fig.~\ref{fig:Zpt}, where we show the value of $T(\lambda)$  associated with the fiducial cross section obtained imposing $p_{T}^{Z}>p_{T}^c$ and an asymmetric cut on $y_Z$ for several values of the gluon mass $\lambda$. To obtain the functional form of $T(\lambda)$, and hence the slope for $\lambda=0$, we use the following interpolating function 
\begin{equation*}
T_{\rm \scriptscriptstyle fit}(\lambda) = a \left[1 + b \left(\frac{\lambda}{p_T^c}\right)+c\left(\frac{\lambda}{p_T^c}\right)^2\log^2\left(\frac{\lambda}{p_T^c}\right)+d\left(\frac{\lambda}{p_T^c}\right)^2\log\left(\frac{\lambda}{p_T^c}\right) \right].
\end{equation*}
The fitted $b$ coefficient for all the cuts we have considered is always compatible with zero.

\begin{figure}[t!]
\includegraphics[height=0.212\textheight]{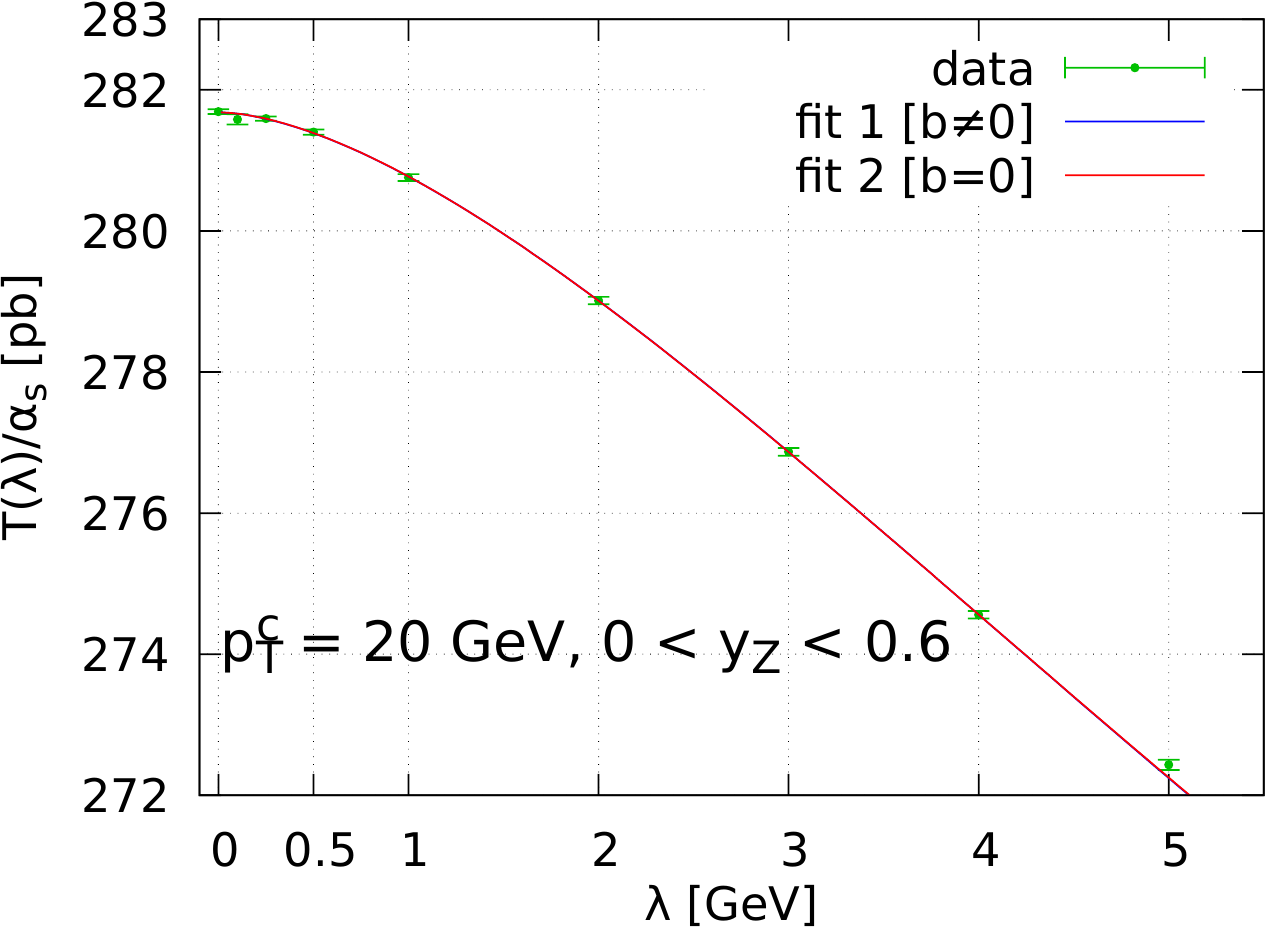}\hfill
\includegraphics[height=0.212\textheight]{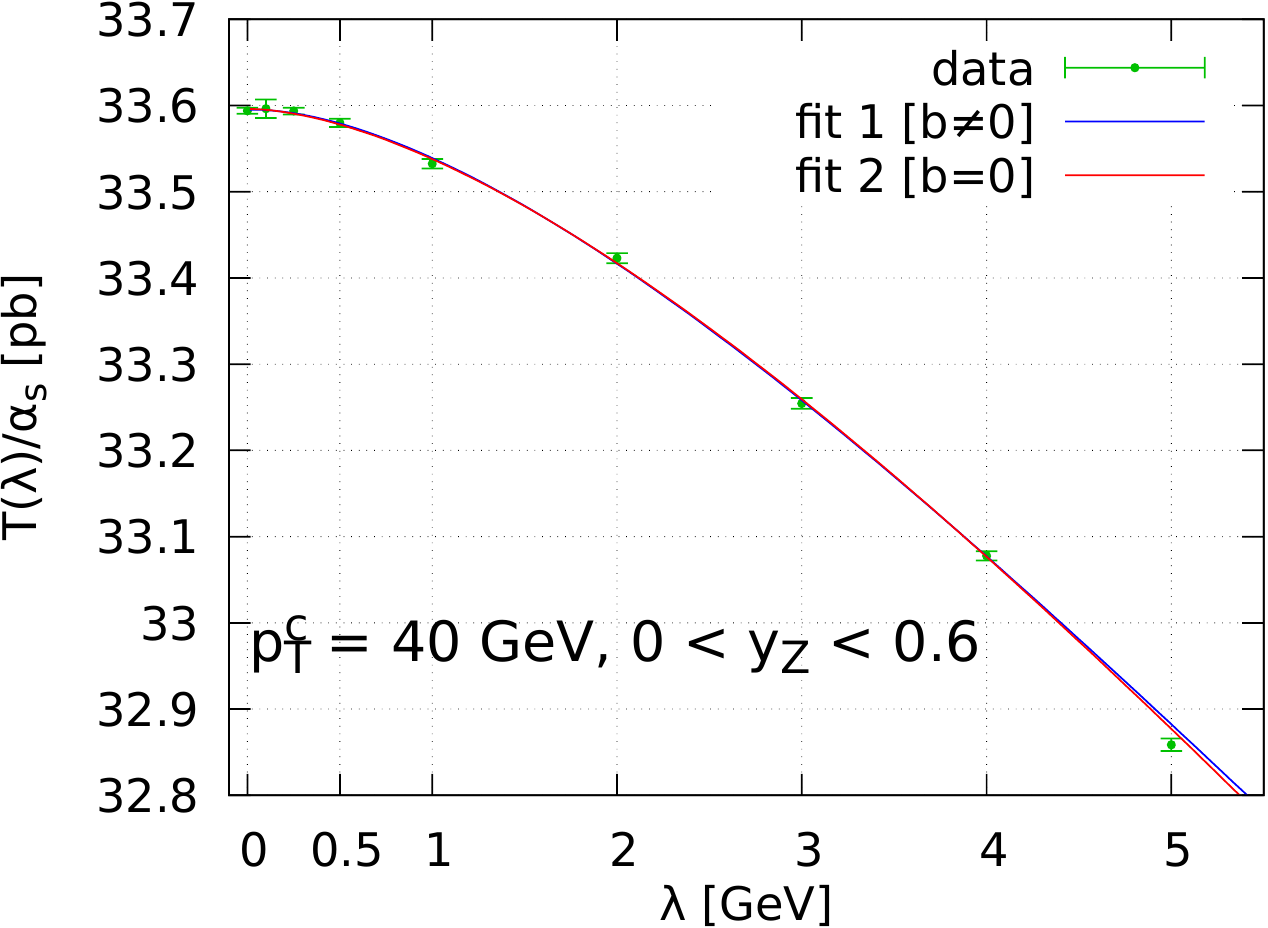}
\caption{$T(\lambda)$ function for the fiducial cross section for $q\gamma\to qZ$ imposing a $Z$-boson rapidity cut $0<y_Z<0.6$ and a transverse momentum of 20~GeV~(left panel) or 40~GeV~(right panel) for a $\sqrt{s}=300$~GeV center-of-mass energy.}
\label{sec:pTZ}
\end{figure}

We can thus conjecture that linear power corrections only arise from soft radiation that, at least in the leading colour approximation, can be emitted only from dipoles. If we integrate over the whole phase space of the emitting dipole, such linear terms cancel.
Further analytical work is needed
in order to put these conjectures on more solid ground~\cite{Caola}.

\section{Shape observables}
\label{sec:shape-obs}

The simplest observables measured at lepton colliders are event shape observables like the thrust
\begin{equation}
\tau = 1- \max_{\vec{n}}  \frac{\sum_{i}|\vec{p}_i. \vec{n}|}{\sum_{i} |\vec{p}_i|},
\end{equation} 
and the $C$-parameter
\begin{equation}
C=3-\frac{3}{2}\sum_{i,j} \frac{(p_i . p_j)^2}{(p_i.Q) (p_j.Q)},
\end{equation}
with $p_i$ the four-momenta of all the final-state particles and $Q=\sum_i p_i$,
as they provide us a continuous measure of deviation from lowest-order process $e^+e^-\to q\bar{q}$ and can serve as a QCD ``laboratory'' to perform, for example, strong coupling-constant fits. 

Non-perturbative corrections are crucial to perform $\alpha_s$ extractions and they can be inferred using a renormalon-inspired dispersive approach. Such corrections are modelled as a negative shift in the perturbative distribution~\cite{Dokshitzer:1997ew}, and they are computed considering the change in the shape obserable induced by a soft emission starting from the two-jet configuration~\cite{Dokshitzer:1995zt}, where the strong coupling is replaced with an infrared-regular effective strong coupling similar to the one in eq.~\eqref{eq:effcoupl}.
 The authors of Ref.~\cite{Nason:1995np} investigated the effect of infrared renormalons upon the thrust using the large-$n_f$ limit, showing the additional branching
of a soft gluon (namely the second line of eq.~\eqref{eq:Tlambda}) leads to a larger shift than the naive estimate one would obtain simply considering the effect of a soft emission.
One can however recover the complete result predicted by the large-$n_f$ method by applying a corrective factor, usually dubbed as \emph{Milan factor}~\cite{Dokshitzer:1997iz,Dokshitzer:1998pt,Dasgupta:1999mb}, which is the same for the same class of observables and in particular is identical for the thrust and the $C$-parameter.
In addition to the final $g\to q\bar{q}$ splitting considered by the large-$n_f$ approximation, the inclusion of the Milan factor also accounts explicitly for the final state with two soft gluons, which
gives a sizeable contribution that cannot
 be captured by the naive non-abelianization procedure $n_f \to - 6\pi b_0$, as the two branchings lead to different kinematics.  
Alternatively, one could use Monte Carlo~(MC) event generators to estimate such non-perturbative corrections~\cite{Bethke:2008hf,Dissertori:2009ik,OPAL:2011aa,Kardos:2018kqj}. This approach has however been criticised since the separation between perturbative and non-perturbative components in a MC is rather arbitrary. Furthermore, the hadronization model is tuned on parton showers which provide a less accurate description of the perturbative component.

Current $\alpha_s$ extractions from shape observables, where hadronization corrections are extracted from a renormalon-inspired analytic model, are roughly three standard deviations smaller the world average $\alpha_s(M_{\small Z})=0.1179(10)$~\cite{pdg}, for example
\begin{itemize}
\item $\alpha_s(M_{\small Z})= 0.1135(10)$ from the thrust distribution~\cite{Abbate:2010xh};
\item $\alpha_s(M_{\small Z})= 0.1123(15)$ from the $C$-parameter distribution~\cite{Hoang:2015hka}.
\end{itemize}
The authors of Ref.~\cite{Luisoni:2020efy} have suggested that this could be due to the fact that measurements are usually performed away from the two-jet limit, where the non-perturbative shifts $\delta \tau$ and $\delta C$ are estimated.
Indeed the shift in the cumulant from the $C$-parameter distribution
\begin{equation}
\bar{\Sigma}(C) = \int_C^{1} dC^\prime \frac{d\sigma}{dC^\prime} d C^\prime \rightarrow \bar{\Sigma}(C) - \delta C \frac{d\sigma}{dC}
\label{eq:cumulant}
\end{equation}
computed for $C=3/4$, which corresponds to the three-jet symmetric point, is roughly half of the value for $C=0$
\begin{equation}
\delta C(3/4) = \frac{C_A + 2 C_F}{2C_F} \times 0.224 \, \delta C(0) = 0.48\, \delta C(0).
\label{eq:deltaC3}
\end{equation}
This number is obtained by noticing that there is a Sudakov shoulder at $C=3/4$, so one could estimate non perturbative corrections in the same fashion as done for the singular $C=0$ point.
As noted by the authors of Ref.~\cite{Luisoni:2020efy}, that procedure is able to give a rigorous prediction only at the singular configurations, $C = 0, 3/4$, not at other points of the spectrum, because different recoil prescriptions employed to model the kinematic of soft emissions can lead to different linear terms. 

\begin{figure}[t!]
\includegraphics[width=0.5\textwidth]{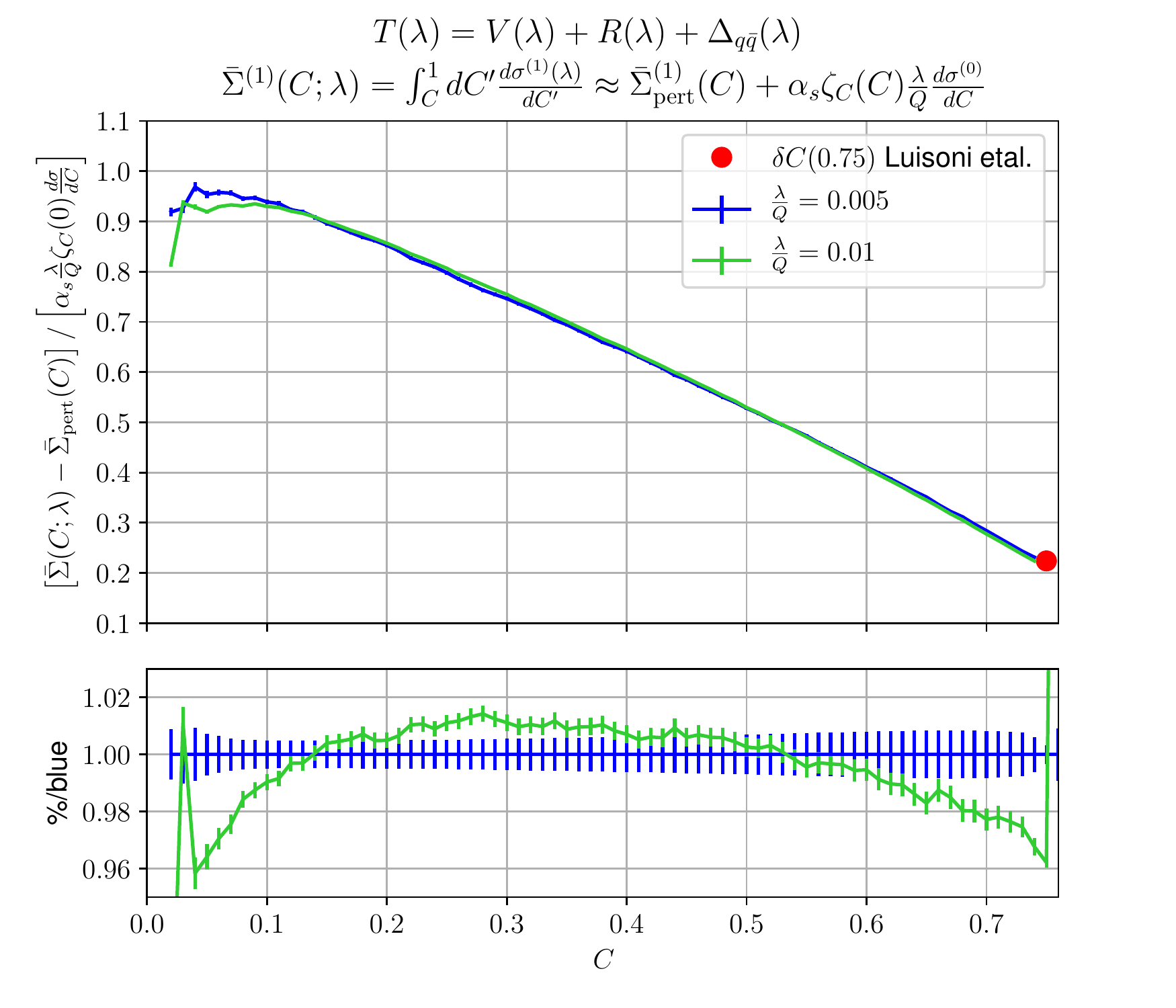}
\includegraphics[width=0.5\textwidth,page=2]{pythonPlots}
\caption{
Non-perturbative shift in the cumulant distribution for the $C$ parameter~(left) and the thrust contribution~(right) computed using $\lambda=\frac{1}{100}Q$~(green) and $\lambda=\frac{1}{200}Q$~(blue) for the LO process $e^+e^-\to q\bar{q}\gamma$.
The red line in the top-left panel corresponds to the relative shift in eq.~\eqref{eq:deltaC3}, computed in Ref.~\cite{Luisoni:2020efy}, obtained setting $C_A=0$.
 }
\label{fig:cumulants}
\end{figure}

We can however extract the dependence of the shift on the geometry of the event in the large-$n_f$ limit, allowing us to investigate also configurations and shape observables which do not feature additional Sudakov shoulders. Since this abelian method works only if there are no gluons at the lowest perturbative order, to move away from the two-jet limit we can consider the $e^+e^- \to q\bar{q}\gamma$ configuration as LO process~\cite{Caola}.
QED divergencies are subtracted using the same procedure employed in Ref.~\cite{FerrarioRavasio:2020guj}.
We thus compute the $T(\lambda)$ function for the integrated cross section with a lower cut on $C$ or $\tau$ for a small value of $\lambda$, and then subtract $T(0)$, which corresponds to the standard $\mathcal{O}(\alpha_s)$ correction to the LO prediction.
Form eq.~\eqref{eq:cumulant} we see that to extract the non-perturbative shift we can just divide the difference $T(\lambda)-T(0)$ by the LO differential distribution computed for the 3-jet configuration $e^+e^-\to q \bar{q} \gamma$.
The result, normalized by the value of $\lambda$ and by the non-perturbative shift computed in the two-jet limit\footnote{This is the shift in the $T(\lambda)$ function, that needs to be convoluted with an effective coupling to get the full result, and thus is different from the shifts in eq.~\eqref{eq:deltaC3}. However this is not necessary if we are just interested in assessing the dependence on the geometry of the event, \emph{i.e.} $\delta C(C)/\delta C(0)$ for $C>0$.}
\begin{equation}
\delta C(0) = \frac{15 \pi^2}{16}, \quad \delta \tau(0) = \frac{2\pi}{3}\delta C(0),
\label{eq:shift0}
\end{equation}
is shown in the upper panels of Fig.~\ref{fig:cumulants} for the $C$-parameter~(left) and the thrust~(right). 
From the bottom panels we see that the curves computed with different small values of $\lambda$ agree very well between each other,
and the differences are compatible with additional suppressed quadratic power corrections.
We also observe that, by reducing the value of $\lambda$, the shift computed for a low cut on $C$ or $\tau$ converges to the values in eq.~\eqref{eq:shift0}.
 From the left panel of Fig.~\ref{fig:cumulants} we notice that the shift in $C=3/4$ is consistent with $0.224 \delta C(0)$, which corresponds to the value found in Ref.~\cite{Luisoni:2020efy} for $C_A=0$.
We stress that we do not need a Sudakov shoulder to predict the linear-$\lambda$ sensitivity, so we are also able to compute $\delta \tau$ for any values of $\tau \le 1/3$.
These findings support the issue raised in Ref.~\cite{Luisoni:2020efy} that non-perturbative corrections in shape observables depend on the geometry of the event. 

\section{Conclusion and Outlooks}
\label{sec:concl}
In the present contribution we reviewed the basic properties of linear power-corrections $\Lambda/Q$ affecting kinematic distributions at colliders. 
Power corrections originate from the bad large-order behaviour of QCD, and signal the incompleteness of perturbative expansions and the need for the inclusion of non-perturbative corrections to recover the full result.
For the majority of the kinematic distributions measured at LEP and at the LHC there is no OPE, so one needs to use approximations to perform all-orders calculations in QCD to infer the large-order behaviour and hence the non-perurbative corrections.
All-order corrections become accessible in the large-$n_f$ limit, with $n_f$ being a fictitious number of light flavours. To recover the non-abelian property of QCD, at the end of the calculation one performs the $n_f \to -6\pi b_0$ replacement. Since infared renormalons are mainly caused by the divergence of the strong coupling at small scales, this procedure, which encapsulates an explicit dependence of the running of the strong coupling via the $b_0$ coefficient, provides us a simple but yet powerful way to assess the presence of linear $\Lambda$ corrections.
In particular in this contribution we focused on some recent phenomenological applications of this method.
\begin{itemize}
\item Single-top production and decay~\cite{FerrarioRavasio:2018ubr}. Inclusive observables that do not depend upon the kinematics of the coloured final states, like the total cross section or leptonic observables, should be computed in the ${\rm \overline{MS}}$ scheme (or in any other short distance mass scheme). However, if one employs the NWA, leptonic observables will still be affected by linear renormalons.
If we instead want to calculate the mass of the top decay products, the pole mass yields a better perturbative convergence as it allows for a partial cancellation between the pole-mass renormalon and the physical one present in the intrinsic definition of the observable. This cancellation is exact in the limit of a vanishing top-width and if we are inclusive enough in the definition of the $b$-jet arising from the top decay.
Indeed, if we define the $b$-jet using a small jet radius $R$, we are strongly affected by
jet-related  linear corrections, which inversely proportional to $R$, irrespectively of the mass prescription adopted.
\item The transverse momentum of the $Z$-boson produced in association with a hard jet in hadronic collisions~\cite{FerrarioRavasio:2020guj}. For this observable we do not find any numerical evidence of a linear sensitivity despite the absence of an azimuthally-symmetric radiation pattern. This finding is not altered by the imposition of additional asymmetric rapidity cuts.
\item Shape observables, where the estimate of hadronization effects is important to perform reliable extractions of the strong coupling.
The authors of Ref.~\cite{Luisoni:2020efy} have raised the issues that measurements are performed away from the two-jet limit, where non-perturbative corrections are inferred, considering explicitly the case of $C=3/4$.
By studying the process $e^+e^-\to q\bar{q}\gamma$, we are able to confirm their finding and make predictions for any shape-observable, however without the inclusion of non-perturbative corrections proportional to $C_A$~\cite{Caola}.
\end{itemize}
Further analytical work is needed
in order to formulate an analytic argument to explain the cancellation of linear power-corrections for processes where the soft-radiation pattern is not symmetric. 
This would also allow a major understanding of the underlying mechanisms that lead to the presence or absence of renormalon effects in  collider physics.
Sec.~\ref{sec:shape-obs} contains preliminary results that will be published in Ref.~\cite{Caola}.

\paragraph{Acknowledgements} 
I want to thank Irinel Caprini and Diogo Rodrigues Boito for the invitation to contribute to the special issue ``Renormalons and Hyperasymptotics in
QCD''.
The majority of the material presented in this contribution is obtained in collaboration with Fabrizio Caola, Giovanni Limatola, Kirill Melnikov, Paolo Nason and Carlo Oleari, to whom I am very grateful.
I also want to acknowledge Pier F. Monni and Gavin P. Salam for useful discussions and comments on the manuscript.
This work is supported by the
European Research Council (ERC) under the European Union’s Horizon 2020 research and innovation program (grant agreement No. 788223, PanScales).

\end{document}